%% file: main.tex
\documentclass[twocolumn,groupedaddress]{aastex63}

\usepackage{amsmath}
\usepackage{empheq}
\usepackage{mathrsfs}
\usepackage{textcomp}
\usepackage{hyperref}
\usepackage{graphicx}
\usepackage{xspace}
\usepackage[caption=false]{subfig}
\usepackage{multirow}
\usepackage{longtable}
\hypersetup{colorlinks, linkcolor={blue}, citecolor={blue}, urlcolor={blue}}

\newcommand{\tobs}{$\theta_{\rm obs}$\xspace}
\newcommand{\costh}{cos$(\theta_{\rm obs})$}
\newcommand{\ejm}{M$_{\rm ej}$\xspace}
\newcommand{\kmsmpc}{km\,${\rm s}^{-1}\,{\rm Mpc}^{-1}$\xspace}

\begin{document}
\title{Constraining the observer angle of the kilonova AT2017gfo associated with GW170817: Implications for the Hubble constant}
\correspondingauthor{Suhail Dhawan}
\email{suhail.dhawan@fysik.su.se}
\author{S. Dhawan, M. Bulla, A. Goobar, A. Sagu\'es Carracedo, C. N. Setzer}
\affiliation{The Oskar Klein Centre for Cosmoparticle Physics, Department of Physics, Stockholm University, AlbaNova, 10691 Stockholm, Sweden}

\begin{abstract} 
There is a strong degeneracy between the luminosity distance ($D_L$) and the observer viewing angle ($\theta_{\rm obs}$; hereafter viewing angle) of the gravitational wave (GW) source with an electromagnetic counterpart, GW170817. Here, for the first time, we present independent constraints on $\theta_{\rm obs} = 32.5 ^{\circ +11.7}_{-9.7}$ from broad-band  photometry of the kilonova (kN) AT2017gfo associated with GW170817.  These constraints are consistent with independent results  presented in the literature using the associated gamma ray burst GRB170817A. Combining the constraints on $\theta_{\rm obs}$ with the GW data, we find an improvement of 24$\%$ on $H_0$. The observer angle constraints are insensitive to other model parameters, e.g. the ejecta mass, half-opening angle of the lanthanide-rich region and the temperature. A broad wavelength coverage extending to the near infrared is helpful to robustly constrain $\theta_{\rm obs}$. While the improvement on $H_0$ presented here is smaller than the one from high angular resolution imaging of the radio counterpart of GW170817, kN observations are significantly more feasible at the typical distances of such events from current and future LIGO-Virgo Collaboration observing runs ($D_L \sim 100$ Mpc). Our results are insensitive to the assumption on the peculiar velocity of the kN host galaxy.
\end{abstract}
\keywords{gravitational waves}
\section{Introduction}
The first detection of gravitational waves (GWs) from a binary neutron star \citep[BNS;][]{2017PhRvL.119p1101A} merger event revolutionised our understanding of the physics of compact objects. The event was detected on 2017 August 17 at 12:41:02 UTC by the Advanced LIGO/Virgo collaboration (LVC) detectors, localised to $\sim 28\, {\rm deg}^2$ with an estimated luminosity distance, $D_L$, of $\sim 40$ Mpc \citep[][hereafter A17:H0]{2017Natur.551...85A}. Follow-up multi-wavelength observations of the LIGO/Virgo sky map led to several independent detections of an electromagnetic (EM) counterpart, associated with the galaxy NGC 4993 \citep{2017Natur.551...64A, 2017Sci...358.1556C,2017ApJ...848L..17C, 2017Sci...358.1570D, 2017Sci...358.1559K,2017ApJ...848L..16S}.

The luminosity distance to BNS events can be determined directly from the GW signal, dubbing these events as GW ``standard sirens" \citep[GWSS;][]{2005ApJ...629...15H}, the GW equivalent of ``standard candles" in the EM spectrum.
Combining this luminosity distance from the GW signal with the redshift to the host galaxy, using information on the position from the EM counterpart, has long been proposed as an effective method to measure cosmological parameters, particularly the Hubble constant \citep[$H_0$; e.g.][]{2013arXiv1307.2638N}. Studies have shown that 20 - 50 events in the redshift range $z \sim 0.1$ can measure $H_0$ at the 1-2$\%$ level \citep{2017ApJ...840...88C,2018arXiv181111723M}.
This is especially interesting since the most precise local distance ladder estimates \citep{2019arXiv190805625R,2019ApJ...876...85R} are in $\gtrsim 4 \sigma$ tension with the inferred value from the cosmic microwave background \citep[CMB;][]{2018arXiv180706209P}. Time-delay distances to strongly lensed quasars also suggest a high local $H_0$, further exacerbating the tension with the CMB inference to $\gtrsim 5 \sigma$ \citep{2019arXiv190704869W}. A summary of the current status of the Hubble tension is provided in \citet{2019arXiv190710625V}. This tension could indicate the presence of exotic physics beyond the standard model \citep[for e.g., see][]{2017JCAP...10..020R,2018JCAP...09..025M,2018JCAP...11..014D,2019arXiv190200534K}. GWSS events present an excellent route for determining $H_0$ and resolving the tension \citep{2019PhRvL.122f1105F}. 

The largest uncertainty in the $H_0$ estimation from GW170817 comes from the degeneracy between $D_L$ and the inclination angle inferred from the GW signal. The inclination angle ($i$) is defined as the angle between the total angular momentum of the binary system and the line of sight from source to earth (which, for the case here of no orbital precession, is related to the viewing angle as, $i = 180 - \theta_{\rm obs}$). Independent information on the binary inclination from the EM signal can be used to improve the constraints on $D_L$ and hence, $H_0$ \citep[as shown for this event, for e.g., in][]{2017ApJ...851L..36G,2019NatAs.tmp..385H}. While these constraints are based on the radio and X-ray emission from the gamma-ray burst (GRB) associated with the GW event, here we constrain the inclination angle using broad-band photometry of the kilonova \citep[kN;][]{2017LRR....20....3M}, AT2017gfo, associated with the GW event. A kN is a transient powered by radioactive decay of $r$-process elements produced in the BNS merger ejecta, which dominates the ultraviolet (UV) to near infrared (NIR) emission of the EM counterpart. Recent studies have proposed the use of the EM signal from the kN for distance measurements and hence, calculating $H_0$ \citep[e.g., see][]{2019arXiv190800889C}. Here, we only focus on the constraints from the EM counterpart on the viewing angle, and its impact on the inferred $D_L$ and $H_0$.

In this study, we constrain the viewing angle of GW170817 with detailed 3-D radiative transfer models using the \texttt{POSSIS} code \citep[hereafter B19:kN;][]{2019MNRAS.489.5037B} and evaluate its impact on the inferred $H_0$. In Section~\ref{sec:data} we present the input datasets and methodology for inferring the viewing angle. In section~\ref{sec:results}, we present our results and discuss and conclude in section~\ref{sec:disc}.

\begin{figure*}
    \centering
    \includegraphics[width=\textwidth]{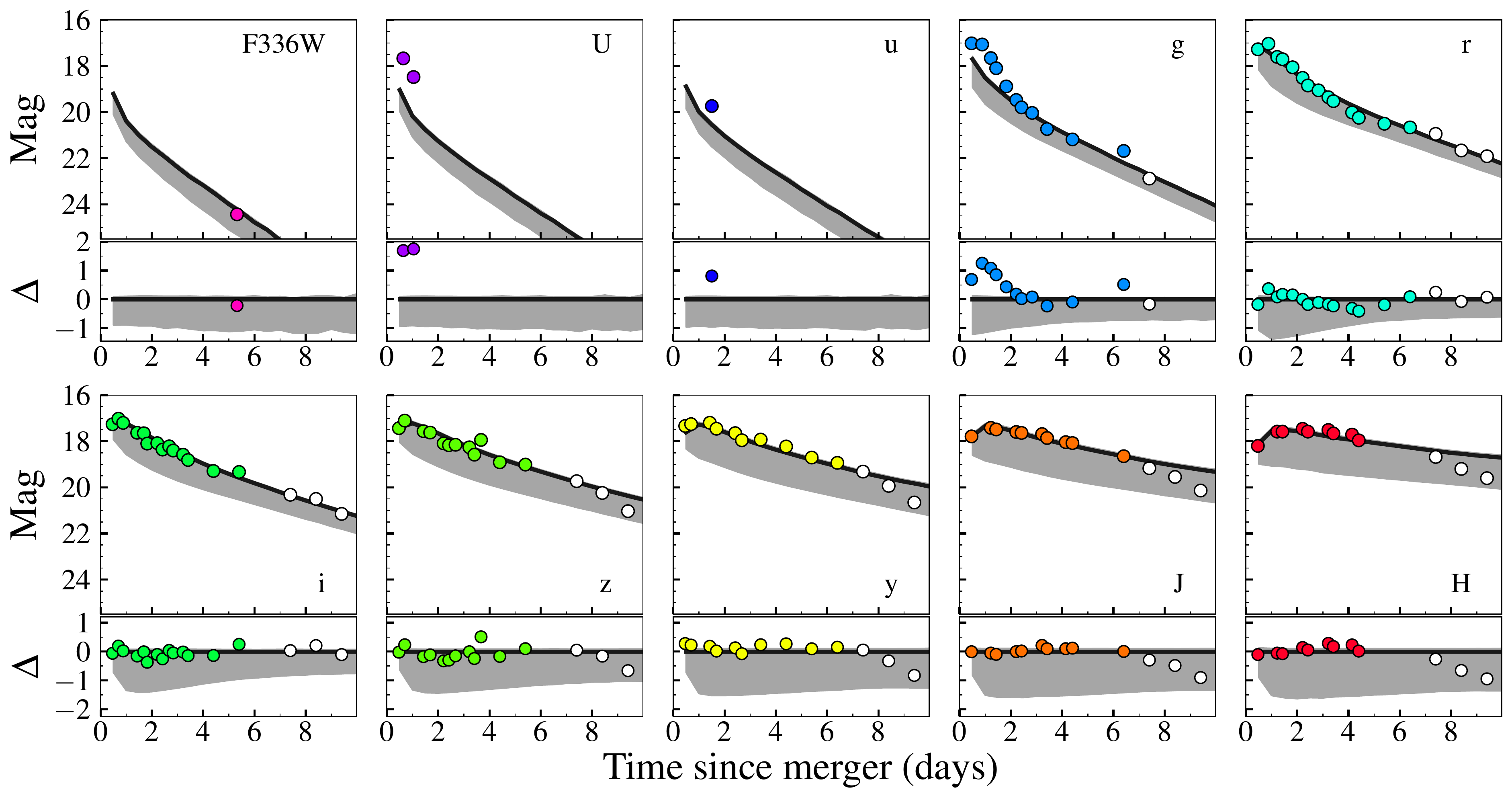}
    \caption{UV to NIR photometry of GW170817/AT2017gfo (circles, corrected for Milky Way reddening) together with our best-fit model with $\cos\theta_\mathrm{obs}=0.9$, $M_{\rm ej}=\,0.05 M_{\odot}$, $\Phi=\, 30^{\circ}$ and $T=5000$~K (black solid lines). Grey shaded areas in each panel mark the range spanned by the same model for different viewing angles (from pole, upper edge, to equator, lower edge). Lower panels show residuals from the best-fit model. Data after 7~d (open circles) are excluded from our fiducial model fit (see Section~\ref{sec:datamodels}), while their impact discussed in Section~\ref{sec:datasets}. }
    \label{fig:bestfit}
\end{figure*}

\section{Data and Methodology}
\label{sec:data}

\subsection{GW data}
For the GW data, we use the joint posterior distribution on ${\rm cos}(i)$ and $H_0$ as reported in A17:H0. For our analyses, we use their fiducial case for $H_0$, which assumes a recession velocity of 3017 $\pm 166$\, km s$^{-1}$ for the host galaxy NGC 4993. Since NGC 4993 is in the galaxy group ESO-508, this recession velocity includes a correction of 310 km s$^{-1}$, accounting for the group velocity  \citep{2014MNRAS.445.2677S,2015MNRAS.450..317C}. The estimated uncertainty includes a conservative value of 150 km s$^{-1}$ for the uncertainty on the peculiar velocity at the location of NGC 4993. A17:H0 show that the impact of the recession velocity is small in the final $H_0$ inference. Moreover, since the aim of this analysis is to test the impact of independent constraints of the viewing angle on the inferred $H_0$, and is focussed on the difference between the GW-only and GW+EM information cases, we have the same assumptions on the recession velocity for both cases. Therefore, our conclusions are independent of the assumptions for deriving the recession velocity.

For our analysis, we use the publicly reported posterior distribution from A17:H0. We do this to be consistent with the inference from the GW-only data, and hence, focus on  the role of the EM prior, by quantifying the difference in the inferred $H_0$.

\subsection{EM data and models}
\label{sec:datamodels}
For our analyses, we use ultraviolet (UV), optical and near infrared (NIR) photometry of GW170817, ranging from $u$ to $H$ band from \citet{2018MNRAS.480.3871C} where the authors have analysed the photometry from \citep{2017PASA...34...69A,2017Natur.551...64A,2017ApJ...848L..19C,2017ApJ...848L..17C,2017Sci...358.1570D,2017Sci...358.1565E,2017Sci...358.1559K,2017Natur.551...67P,2017Natur.551...75S,2017ApJ...848L..27T,2017Natur.551...71T}. We use the $ugrizyJH$ (hereafter, UVOIR) photometry from these studies. In this work, we use models from B19:kN. These models explain the observations of GW170817/AT2017gfo in the first week after the merger, but not at later epochs (likely due to incorrect opacities at those later phases, see discussion in section~4.2 of B19:kN). Therefore, we focus on data up to 7\,d after the merger. The photometry is corrected for extinction due to Milky Way (MW) dust, using the standard MW dust law \citep{1989ApJ...345..245C} with total-to-selective absorption, $R_V = 3.1$ and $E(B-V) = 0.11$ mag from the dust maps of \citep{2011ApJ...737..103S}. We do not correct for host galaxy extinction since it is expected to be low \citep[see, e.g.][]{2017Natur.551...67P}. We test the impact of this assumption in section~\ref{sec:datasets}.

We analyse the data with the Monte Carlo (MC) radiative transfer software \texttt{POSSIS} (B19:kN), a code that calculates synthetic observables, e.g. spectra, light curves and polarisation for transient events, e.g. supernovae and kNe. \texttt{POSSIS} is well-suited to study 3-D ejecta geometries and thus predict observables at different viewing angles. In this work, we adopt the kN model which is characterized by a ``lanthanide-rich" component around the equator and a ``lanthanide-poor" component at higher latitudes. Several studies in the literature show that simulations consisting of a two-component model are an appropriate descriptions of kNe \citep[e.g;][]{2013ApJ...773...78B,2014MNRAS.441.3444M,2017PhRvD..95h3005S,2017PhRvL.119w1102S}. Moreover, a two-component model has been shown to explain the observations of GW170817 well \citep{2017Natur.551...67P,2017Natur.551...75S,2017ApJ...848L..17C,2019NatAs...3...99B}.

\texttt{POSSIS} uses a parametrised form of the opacity computed from numerical simulations in \citet{2018ApJ...852..109T}. These simulations compute the opacity for the lanthanide-free
region assuming a high-electron fraction, $Y_{\rm e}$ of 0.3. For the lanthanide-rich region, the authors do not use a single value of $Y_{\rm e}$ to compute the opacity, but instead use a flat distribution of $Y_{\rm e}$ from 0.1 to 0.4, that is found to reproduce nucleosynthetic yields in agreement with the solar abundance of $r$-process elements \citep{2004ApJ...617.1091S}. Therefore, the assumption on the opacity of the lanthanide-rich region is not based on a single value of $Y_{\rm e}$ but instead an ensemble average computed to match the observed abundance ratio of $r$-process elements. We discuss the impact of this assumption on our results in section~\ref{sec:disc}.

Observables predicted by \texttt{POSSIS} for the two-component kN model depend on three main parameters: the total ejecta mass, \ejm, the half opening angle of the lanthanide-rich component, $\Phi$, and the temperature of the ejecta at 1~day after the merger, $T$. In B19:kN, the temperature was fixed to $T=5000$~K and a good fit to broad-band photometry of GW170817 was found for a model with $M_{\rm ej}=\,0.04 M_{\odot}$ and $\Phi=\, 30^{\circ}$. As summarized in Table~\ref{tab:model_param}, here we extract observables for a grid of 200 models, in which \ejm\, is allowed to vary between 0.01 to 0.10~$M_{\odot}$, $\Phi$ between 15 and 75~$^{\circ}$ and $T$ between 3000 and 9000~K. The computed synthetic observables are then marginalised over \ejm, $\Phi$ and $T$ to obtain a distribution of allowed values of the viewing angle. Here, we find the best fit value of $M_{\rm ej}=\,0.05 M_{\odot}$ and $\Phi=\, 30^{\circ}$ owing to a higher resolution of the model grid computed compared to B19:kN (see Table~\ref{tab:model_param}).
We discuss the implications of marginalising over these ejecta parameters on the inferred \tobs\, probability distribution.  We note here that models with higher M$_{\rm ej}$ predict brighter luminosities than models with lower M$_{\rm ej}$. Similarly, models with low $\Phi$ values (i.e. a smaller half opening angle of the lanthanide-rich region) are brighter than models with larger $\Phi$. This is also true for models with viewing angle closer to the polar region.
We find that the models predict the observed brightness well, except in the $u$-band and the early data in the $g$-band. This is due to the assumptions in computing the wavelength dependence of the opacity. We discuss the impact on the inferred \tobs\, distribution below.
\begin{figure*}
    \centering
    \includegraphics[width=.8\textwidth]{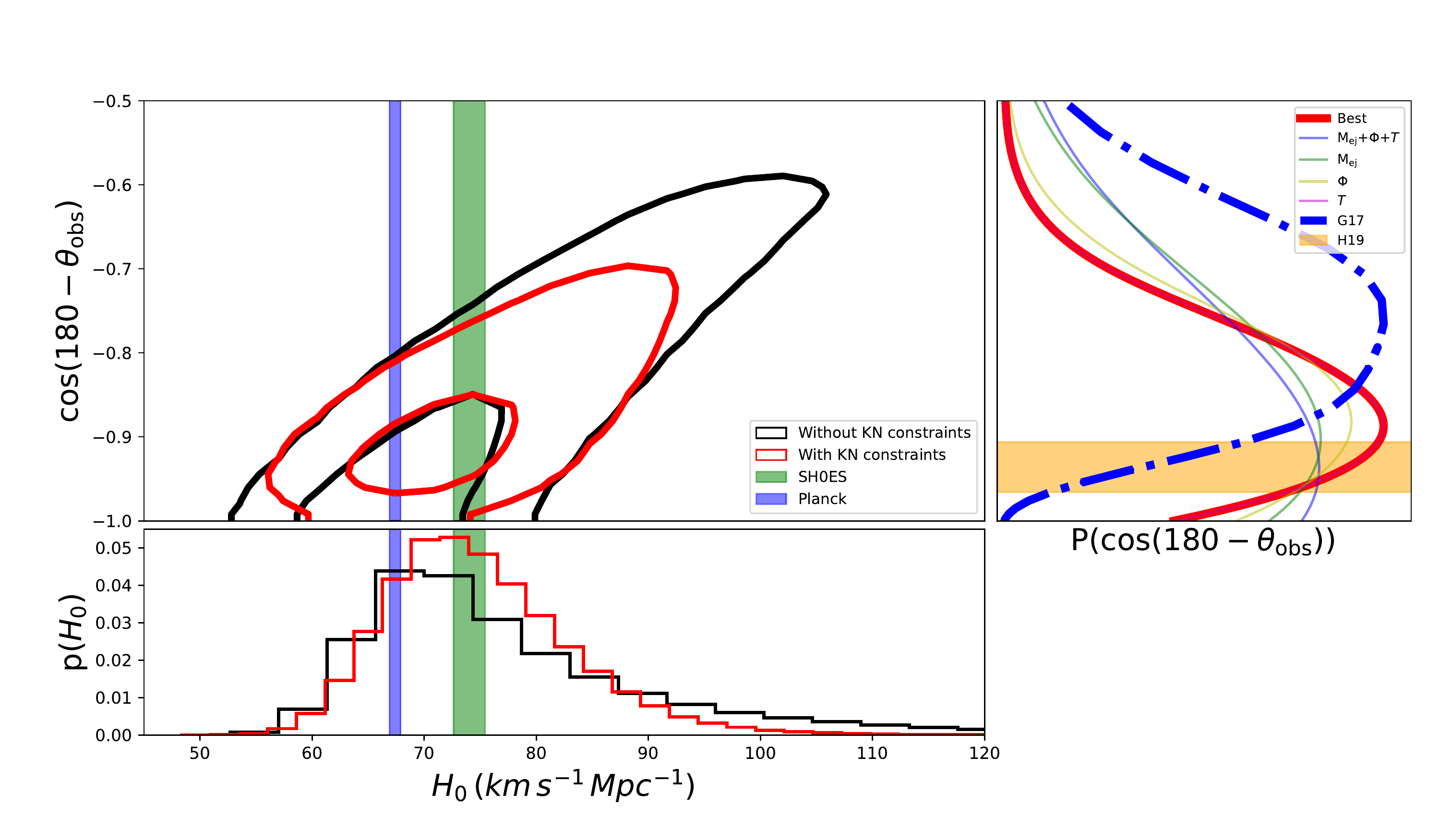}
    \caption{(Left): The 2-D posterior distribution of $H_0$ and cos(180-$\theta_{\rm obs}$) from the LVC GW data \citep[black;][]{2017Natur.551...85A}. Independent constraints on the inclination lead to a narrower posterior distribution with a slightly shifted median value (red). The SH0ES  \citep{2019ApJ...876...85R} and \emph{Planck} \citep{2018arXiv180706209P} $H_0$ values along with the 1-$\sigma$ region are plotted for comparison. (Right): The prior on \costh\, from the kN photometry with the model parameters fixed to the best fit values from B19:kN (red) along with the case with  marginalising over (blue), ejecta mass (green), half opening angle of the lanthanide-rich region (yellow) and temperature (magenta). The \costh\, constraints  from \citet{2019NatAs.tmp..385H} are shown as the orange shaded region  and from \citet{2017ApJ...851L..36G} as the blue dash-dotted line for comparison. (Bottom): The marginalised 1-D posterior distribution of $H_0$.}
    \label{fig:h0_dist}
\end{figure*}
\section{Results}
\label{sec:results}

\input{model_parameters.tex}
In this section, we present the resulting $H_0$ distribution including constraints on the viewing angle from 3-D modelling of the broad-band kN photometry using \texttt{POSSIS} (B19:kN).

We fit the broad-band synthetic photometry to the data described in Section~\ref{sec:data} (see  Figure~\ref{fig:bestfit} for the model fit and  Table~\ref{tab:model_param} for the resulting parameters) to obtain the 1-D probability distribution for \costh. The resulting prior distribution for ${\rm cos}(180 - \theta_{\rm obs})$, is shown in Figure~\ref{fig:h0_dist}.

We smooth the prior distribution for \costh using a cubic spline interpolation. In A17:H0 the authors use a uniform prior on the inclination and not on \costh. We, therefore, reweight the probability distribution by the prior on \costh~used in the GW analysis (see Figure~\ref{fig:h0_dist}). 
Accounting for both the prior from the GW analysis and the kN photometry (so as not to reweight the posterior distribution twice), we obtain the $H_0$ distribution plotted in Figure~\ref{fig:h0_dist}. The resulting value of $H_0$ is  72.4$^{+7.9}_{-7.3}$ \kmsmpc, compared to the value of 70.0$^{+12.2}_{-7.8}$ \kmsmpc from the GW-only data in A17:H0. We note that this number is computed from a spline interpolation of the LIGO posterior distribution. The 68$\%$ maximum a posterior (MAP) region is 24$\%$ smaller than the MAP for the $H_0$ inferred from the GW-only data.

\subsection{Role of model parameters}
We analyse the impact of the kN model parameters in the inferred value of $H_0$. For our fiducial case, we use the best fit parameters from fitting the model grid to the observations. 
We then test the impact of marginalising over the entire range of model parameters, as detailed in Table~\ref{tab:model_param}. We emphasize that marginalising over the model parameters (\ejm, $\Phi$, $T$) only affects the $H_0$ value via the prior on the viewing angle.
We find $H_0$ of 70.3$^{+11.9}_{-6.3}$ \kmsmpc an improvement in the 68$\%$ MAP region of 11$\%$. 

We also marginalise over each of the individual model parameters keeping the others fixed to their best fit value. After marginalising over \ejm, we find an $H_0$ of 71.1$^{+11.8}_{-6.4}$ \kmsmpc, which is also an 11$\%$ improvement. For $\Phi$, we get $H_0$ of 72.3$^{+9.1}_{-6.8}$ and for T of 72.4$^{+8.3}_{-6.8}$ \kmsmpc. Hence, the largest increase in the uncertainty is due to the correlation between the \ejm and \costh. Therefore, it is important to have a robust determination of the \ejm using 3-D models, compared to two-component 1-D models (see discussion in B19:kN).  The cases with $\Phi$ and T have a similar improvement compared to the fiducial case. For each of the cases tested here, the shift in the central value of $H_0$ relative to the fiducial case (shown in red in Figure~\ref{fig:h0_dist}) is significantly smaller than the 68$\%$ MAP region (see Table~\ref{tab:results}). 

\subsection{Role of individual datasets}
\label{sec:datasets}
We analyse the impact of different subsets of the kN photometry in estimating the \costh~distribution. As described above, we use all the photometry from the $u$ to $H$ filters up to 7 days from the merger as our fiducial case. Here, we compute the \costh\, and $H_0$ distribution  using only a subset of the photometry.  Without using the UV data, the improvement in $H_0$ is 28$\%$, slightly higher than the fiducial value of 24$\%$. This is because the $u$-band observation is not well fit by the models which underpredict the flux in that wavelength region (see also, B19:kN). Hence, the impact of the poor model fit on the \costh\, distribution is very low. 
Moreover, since we do not correct the data for host galaxy reddening, we test the impact of not using a correction by computing the  \costh\, distribution from only the NIR data, the wavelength regime where extinction from host galaxy dust is expected to be smallest. We find that the resulting $H_0$ distribution has a best fit value consistent with the fiducial case and an improvement of 34$\%$ compared to the GW-only case. The consistency between the NIR-only and the fiducial case provides evidence that the host galaxy reddening probably has a negligible impact on the inferred inclination angle. The improvement is largely due to the increased viewing angle dependence of the models in the NIR compared to the optical (see Figure~\ref{fig:bestfit} and the discussion in B19:kN). This is due to the opacity treatment for the lanthanide-rich and lanthanide-free regions in B19:kN, based on the numerical simulations of \citet{2018ApJ...852..109T}. The model fits, therefore, suggest that NIR follow-up for future kNe will be extremely important for robust estimates of \costh\, and hence, improving the constraint on $H_0$. 

 We also analyse the role of the phase coverage on our constraints, to determine how early we need to discover the optical counterpart. We find that removing the data before +1\,d after the merger leads to an improvement of only 19$\%$ and without data before +2\,d the improvement drops to  12$\%$. Therefore, the early time data are crucial for constraining the kN inclination angle. The reported improvement could be even higher for models with better agreement in the bluer wavebands. We note that if we only use the data in the phase range where the blue component dominates (i.e. +2 to +4 days), there is no significant improvement in the inferred $H_0$.

In our fiducial analysis, we only use data at t $< +7$\,d since the description of the opacity is not valid at later phases. 
Here, we also compute $H_0$ using constraints on \costh\, using all of the photometry described above, including data between +7 and +10\,d and find a value of H$_0 =  72.8^{+8.8}_{-7.3}$ \kmsmpc, consistent with the fiducial case. Hence, excluding the latest phase data does not have a significant impact on the inferred $H_0$.
\input{results_tab.tex}

\section{Discussion and Conclusion}
\label{sec:disc}
In this paper, for the first time, we present constraints on the viewing angle of EM-GW source GW170817/AT2017gfo using broad-band UV to NIR photometry of the kN and quantify its impact on the inferred $H_0$. We find, for our fiducial case, a value of $H_0 = 72.4^{+7.9}_{-7.3}$ \kmsmpc.
We find typical improvements between $\sim$ 10 to $\sim 25 \%$ for different assumptions on the model parameters. The constraints on the observer angle presented here are consistent with previous, distance-independent constraints presented in the literature, e.g. from polarimetry \citep{2017NatAs...1..791C,2019NatAs...3...99B}. The \tobs\, constraints presented here are also consistent with distance-dependent limits \citep[for e.g.][]{2018ApJ...854L..31C,2018ApJ...860L...2F}, as well as other, independent constraints from modeling the radio and X-ray photometry \citep[e.g.][]{2017ApJ...851L..36G} and Very Long Baseline Interferometry (VLBI) data \citep[see orange band in Figure~\ref{fig:h0_dist}]{2019NatAs.tmp..385H}. Our constraints are also consistent with recent modeling of the late phase non-thermal emission of AT2017gfo \citep[see][]{2019arXiv190906393H}.

Recent efforts in the literature have constrained the inclination angle using the properties of the gamma-ray burst (GRB) jet associated with GW170817/AT2017gfo \citep[e.g.][]{2017ApJ...851L..36G,2019NatAs.tmp..385H}. Constraints from radio and X-ray light curves of GRB170817 indicate an $H_0 = 74.0^{+11}_{-7.0}$ \kmsmpc \citep{2017ApJ...851L..36G} for the case assuming $\sigma_{v} = 166\, {\rm km} {\rm s}^{-1}$. This is consistent with the constraints presented here, however,  our fiducial case has a slightly higher improvement in the 68$\%$ MAP of $H_0$.  \citet{2019NatAs.tmp..385H} constrain the inclination of the associated GRB using VLBI data to constrain the superluminal jet motion. They find stringent constraints on $\theta_{\rm obs}$ using the radio light curve and VLBI data, suggesting $0.906 < {\rm cos}(\theta_{\rm obs}) < 0.966$. As a result, they obtain a significantly improved $H_0$ value of 70.3$^{+5.3}_{-5.0}$ \kmsmpc. However, we note that these constraints are conditional on the merger having an observed, associated GRB, which would not be expected for all kN events expected in the future \citep{2018ApJ...857..128J}.  In addition, these radio observations require that the event be in a high density interstellar medium, which is not expected for all events, hence, these observations would not be possible for all kNe discovered in the future. Moreover, GW170817 was a very nearby event, and therefore these observations would be extremely time consuming and challenging at the typical distances ($D_L \sim 100$ Mpc) of the expected discoveries in the LVC third observing run \citep[O3; see e.g.][]{2017arXiv170908079C}. Hence, the constraints on the viewing angle from kN observations, as presented here, would be an excellent complement for future, more distant BNS merger events. Our constraints on $H_0$ are also consistent with $H_0$ derived independently, e.g. from Fundamental Plane (FP) and surface brightness fluctuations (SBF) distances to NGC 4993 \citep{2017ApJ...848L..31H,2018ApJ...854L..31C}. 

We find that the constraints are slightly improved when removing UV observations that aren't well fit by the model. The NIR data have the largest improvement in the 68$\%$ MAP region relative to the case with only GW data. This shows that UV observations are not critical for robust constraints, however, the NIR is important to accurately determine $H_0$. We note that this is due to the underlying assumption about the wavelength dependence of the opacity of the lanthanide-rich and lanthanide-free regions, using a parametrised form of results from numerical simulations from \citet{2018ApJ...852..109T} (see B19:kN for a more detailed discussion). 
It would therefore be interesting to see in the future if improving opacity calculations can improve the model fit to $u$ and $g$-band data, and therefore, further sharpen the constraints on $H_0$. We note that the assumption on the $Y_{\rm e}$ for the lanthanide-rich region in our analyses is an average of values from 0.1 to 0.4, and not a fixed, low value, e.g. $Y_{\rm e} \lesssim 0.2$. However, as discussed above, if the opacity of the lanthanide-rich region is lower (keeping the opacity of the lanthanide-free region, the same), it would make the constraints on the viewing angle less stringent. Conversely, if the opacity is higher, the viewing angle constraints can be more stringent. As discussed above, a single value of $Y_{\rm e}$ for the lanthanide-rich region would not produce nucleosynthetic yields in agreement with the solar abundance of $r$-process elements, which justifies the use of the average over a range of $Y_{\rm e}$.

In our analysis, we use the fiducial value of $D_L$ and $H_0$ from the LVC inference in A17:H0. The LVC analysis uses a $V_{\rm recession}$ of 3017 $\pm 166$\, km\, s$^{-1}$, accounting for the motion of the galaxy group. Studies in the literature have suggested a higher error (250 km\,s$^{-1}$) on the recession velocity \citep{2017ApJ...851L..36G} and/or different prescriptions for obtaining $V_{\rm recession}$  \citep{2017ApJ...848L..31H,2019arXiv190900587H,2019arXiv190908627M,2019arXiv190909609N} to infer $H_0$ different from the fiducial analyses of A17:H0. However, we emphasise that the improvement in $H_0$ demonstrated here is only due to the improvement in the $D_L$, hence, our results regarding the improvement on $H_0$ are not dependent on the prescription for the recession velocity. 

\acknowledgments
We would like to thank Samaya Nissanke, Rahul Biswas, Hiranya Peiris and Daniel Mortlock for helpful discussions. We acknowledge support from the Swedish National Space Agency and the Swedish Research Council, especially through the G.R.E.A.T research environment grant.   

    




\bibliographystyle{mnras}
\bibliography{gwss}
\label{lastpage}
\end{document}

%% file: model_parameters.tex
\begin{table}
    \centering
        \caption{Input parameters for the model grid computed with \texttt{POSSIS} along with the best fit value for each parameter.}

    \begin{tabular}{|c|c|c|c|}
    \hline
      Parameter & Range & Step & Best fit\\
      \hline
      M$_{\rm ej}$ (M$_{\odot}$)  & [0.01, 0.1] & 0.01 & 0.05\\
       T  (K) & [3000, 9000] & 2000 &  5000\\ 
       $\Phi (^{\circ})$ & [15, 75] & 15 & 30\\
       cos$(\theta_{\rm obs})$ & [0, 1] & 0.1 & 0.9\\
       \hline
    \end{tabular}
    \label{tab:model_param}
\end{table}

%% file: results_tab.tex
\begin{table}
    \caption{Summary of the H$_0$ values for the different \texttt{POSSIS} model parameters marginalised over to calculate the synthetic observables.}

    \centering
    \begin{tabular}{|c|c|}
    \hline
    Model fit & H$_0$\\
        & (km s$^{-1}$Mpc$^{-1}$)\\
    \hline
    \hline
Best & 72.4$^{+7.9}_{-7.3}$ \\
M$_{\rm ej}$+$\Phi$+$T$ & 70.3$^{+11.9}_{-6.3}$ \\
M$_{\rm ej}$ & 71.1$^{+11.8}_{-6.4}$ \\
$\Phi$ & 72.3$^{+9.1}_{-6.8}$ \\
$T$ & 72.4$^{+8.3}_{-6.8}$ \\
    \hline
    {\bf GW only} & {\bf 70.0$^{+12.2}_{-7.8}$} \\
    \hline
    \end{tabular}
    \label{tab:results}
\end{table}